# Neutron Guide Building Instruments of the Brazilian Multipurpose Reactor (RMB) Project

A.P.S. Souza,[a,*] L.P. de Oliveira,[a] F. Yokaichiya,[b] F.A. Genezini,[a] and M.K.K.D. Franco[a]

[a] *Projeto do Reator Multipropósito Brasileiro,*
   *Instituto de Pesquisas Energéticas e Nucleares (IPEN), São Paulo – Brazil*

[b] *Departamento de Física,*
   *Universidade Federal do Paraná (UFPR), Curitiba – Brazil*
   *E-mail:* alexandre.souza@ipen.br

ABSTRACT: A growing community of scientists has been using neutrons in the most diverse areas of science. In order to meet the researchers demand in the areas of physics, chemistry, materials sciences, engineering, cultural heritage, biology and earth sciences, the Brazilian Multipurpose Reactor (RMB) will provide 3 thermal guides and 3 cold guides, with the installation of several instruments for materials characterization. In this study, we present a standard design requirement of two primordial instruments, namely Sabiá and Araponga. They are, respectively, cold and thermal neutron instruments and correspond to a Small-Angle Neutron Scattering (SANS) and High-Resolution Powder Neutron Diffractometer (HRPND) to be installed in the Neutron Guide Building (N02) of RMB. To provide adequate flux for both instruments, we propose here an initial investigation of the use of simple and split guides to transport neutron beams to two different instruments on the same guide. For this purpose, we use Monte Carlo simulations utilizing McStas software to check the efficiency of thermal neutron transport for different basic configuration and sources. By considering these results, it is possible to conclude that the split guide configuration is, in most cases, more efficient than cases that use transmitted neutron beams independently of source. We also verify that the employment of different coating indexes for concave and convex surfaces on curved guides is crucial, at least on simulated cases, to optimise neutron flux (intensity and divergence) and diminish facility installation cost.

---
*Corresponding author.

# 1. Introduction

Brazil has four research reactors, where two of them, IEA-R1 ($5\,MW$ pool-type reactor that was built and commissioned in 1957) and IPR-R1($100\,kW$ TRIGA Mark I reactor that was installed in 1960) are appropriate for radioisotope production. Nevertheless, this production is not able to supply national demand in areas as medicine, for instance. The international 99-Mo supply crisis of 2009 aggravated Brazilian nuclear medicine services and consequently encouraged the Federal Government to construct a new research reactor in the country. This facility could create a national strategic infrastructure in nuclear research and also make Brazil a self-sufficient country in radiopharmaceuticals. The sustainability and feasibility of the project were ensured according to recommendations of IAEA. In 2010, an international committee of a bilateral agreement between Brazil and Argentina, namely COBEN (Bi-national Commission on Nuclear Energy), decided to adopt the conceptual model of the Australian research reactor OPAL as a base for the new reactor projects of both countries [1]. The OPAL was designed and built by the Argentinian technological company INVAP [2], which is also responsible for the project of the new Brazilian Multipurpose Reactor (RMB). In these terms, the RMB possesses the Australian reactor as a reference for radioisotope production and instruments that use neutron beams.

The RMB will be an open pool reactor type with a maximum power of $30\,MW$, located at Iperó, a municipality about $100\,km$ from São Paulo city. It will possess heavy water and beryllium as reflectors with light-water moderation. Besides, RMB will use low enriched uranium (19.75%) as fuel, which is planned to be arranged in a $5 \times 5$ matrix core that contains 23 MTR fuel elements.

RMB conceptual project also predicts nuclear fuels and structural materials irradiation testing, post-irradiation analysis and scientific research using neutron beam besides fulfilling Brazilian domestic radiopharmaceuticals demand. The reactor creation will stimulate all science and engineering areas correlated to RMB, which will positively (and consequently) affect the national industry, research, and universities. All expertise and effort to operate RMB and its instruments will certainly push Brazil to the next and advanced level in nuclear technology and development.

As a multipurpose reactor, the RMB project also contains applications in neutron research and instruments. The neutron applications of the RMB project are the irradiation process, which occurs inside the reflector tank, and measurement process, which takes place at the N02 building (number 14 in Figure 1) and is often complementary to synchrotron radiation investigations. According to the basic project, there will be available in-core positions for materials irradiation tests and positions inside reflector for applying the neutron activation analysis (NAA) technique, which possesses numerous applications, e.g., in geology, archaeology, biology, medicine, environment, industry, chemistry, nutrition, and agriculture.

Concerning applications inside N02, the detailed project of RMB contains, according to the main current research reactors, a series of state-of-art instruments that still have to be properly defined. In this study, we present two priority neutron scattering instruments that compose the detailed project of the RMB neutron guide hall (N02), namely Small-Angle Neutron Scattering (SANS) and the High-Resolution Powder Neutron Diffractometer (HRPND).

SANS is a flexible and disseminated technique for studying object structures on the nanoscale. Besides its unique property in investigating magnetic material, not contemplated by traditional SAXS technique (Small Angle X-Ray Scattering). SANS is used in looking to polymer and biological molecules, rock pores, and defects in metal and ceramic materials, for instance. Also, properties of elements measured and inferred with SANS, like length scale, scattering length distribution, and non-destructive are different from those of the X-ray technique, which allows the use of contrast variation method.

In the late nineties, the inauguration of the UVX light synchrotron source, at the Brazilian LNLS laboratory, has brought a large number of scientists to the national X-ray instruments community. This source, which was operating until 2019 August, possess expressive statistics concerning the number of users and researches with published papers. In 2018, for example, there were 566 accepted proposals for external researchers with 212 published papers, where about 17% of users were foreigners, mainly from Argentina [3].

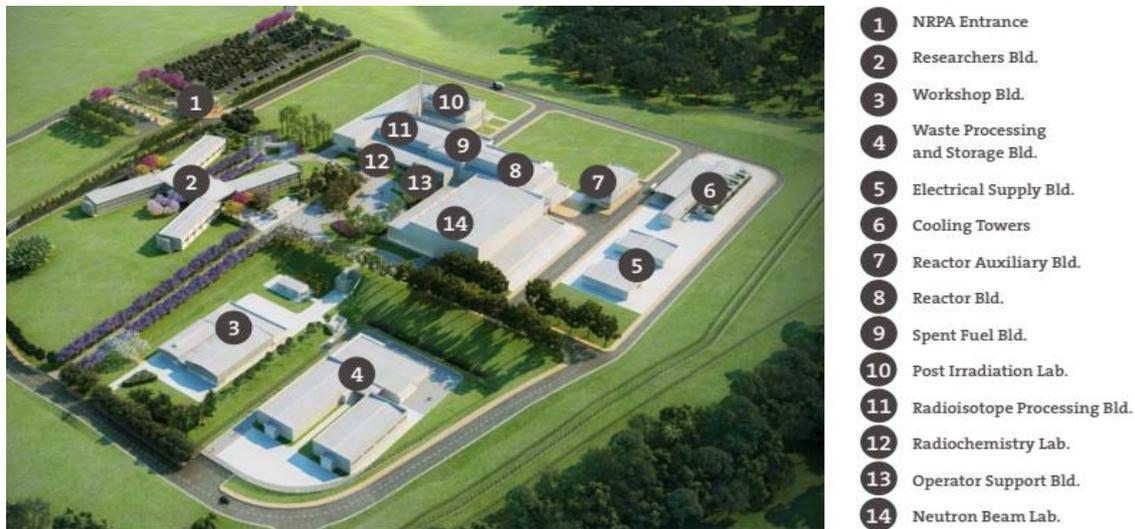

**Figure 1.** Schematic view of RMB Buildings (Bld).

This community of UVX users will probably become larger with the future inauguration of Sirius, a new light synchrotron source of the fourth generation and substitute of the nineties source. The SAXS users are mainly composed of groups that work with condensed matter physics, pharmaceutical sciences, new materials engineering, as well as, e.g., cement and other materials of industry. In this sense, techniques like SANS and neutron diffraction will have a great positive impact on the Brazilian scientific community. RMB SANS instrument was named Sabiá, a symbol bird representative of the Brazilian ornithological fauna and popularly considered the "National Bird of Brazil". SANS-Sabiá will be located at the N02 along the cold neutron guide [4].

Neutron Powder diffraction is a well-established technique for the understanding of the solid state of matter. Similarly to the SANS technique, neutron powder diffraction can study magnetic structures in opposition to X-Ray powder diffraction. This technique allows investigating crystal structures, materials phases, magnetic materials and light elements (and their isotopes) in the presence of heavy ones. A High-resolution powder neutron diffractometer (HRPND) combined with the Rietveld method brings a possibility to distinguish detected Bragg peaks, which helps in determining accurately different phases of powder sample and consequently complex atomic and magnetic structures.

RMB HRPND is named after other native Brazilian bird Araponga (also known as Bare-throated bellbird). Araponga, according to the basic project, will be located at N02 along the thermal neutron guide TG1 [4] and will allow new types of analysis for X-Ray diffraction users and the crystallographic community. Therefore, the powder neutron diffraction of RMB will open possibilities of complimentary analysis for users of Sirius and allow the creation of new groups of research. Considering that, Araponga could fulfil the demand of industry and research in analysing superconductors, alloys, cement, optical materials, minerals, pharmaceuticals and newly created materials.

The definition of instruments and respective guides depends on a large number of parameters and factors, e.g., the desired wavelength range, guide geometry, and neutron profile distribution at the source. The dependence on these factors is so determinant that makes any guide system study practically unique. Generally, all these investigations, which are bound to the reactor (or spallation) source profile, intensity, and geometry, have no validity for any scenario. By considering that, there are some general topics not well described in literature like the efficiency of split guide neutron flux against non-split guide neutron flux.

The use of both types of guides is fundamental for better distribute neutron flux among all neutron guide hall instruments. For example, in OPAL both diffractometers allocated on TG1 share the same guide, where the upstream neutrons, not selected by Wombat monochromator, are conducted to Echidna instrument. On the other hand, there is a neutron transportation system at the FRM-II, which is composed of a set of split guides to share neutron flux throughout instruments [5]. Nevertheless, it is not clear in literature if this kind of choice of using a transmitted flux or a split one is something necessarily bound to source wavelength profile or geometry.

To better understand the dependence of both types of fluxes (guides), we propose a set of Monte Carlo ray-tracing simulations to determine the transmitted and split flux at a specific position of RMB N02 building. These simulations are performed using McStas software, which is constantly updated and supported by DTU Physics, Institut Laue Langevin (ILL), Paul Scherrer Institute (PSI) and the Niels Bohr Institute (NBI) [6]. The use of McStas in investigating guides properties has been widely disseminated in literature as the study of wavelength cutoff based on S-shaped guide geometry, for instance [7].

Proposed simulations are developed using different sources, according to McStas virtual and MCNP source components, to investigate the dependence of the initial wavelength profile. In this first approach, we define some cases with guides and distances based on OPAL thermal guides, i.e., Bunker guides with length of $40\ m$, section area of $30 \times 5\ cm^2$ and final position (of Echidna) about $15\ m$ from the Bunker exit. These cases, however, are just a simplified configuration for the RMB guide system and a complete study for its proper definition is let for future work. This investigation is proceeding based on two different scenarios. In a first moment, simulations are carried out considering symmetric mirror over guide coating surfaces (i.e., all surfaces possess the same reflectivity index). From this approach, new simulations are performed considering different supermirror in curved guides of the neutron transportation system. Once there is no explicit dependence between source (and geometry) and relative efficiency simulated cases, we are allowed to use present results and most efficient geometries in defining RMB guides and instruments, which consequently can save efforts in future simulations.

This paper is organized as follows. In Section 2, we firstly present the neutron transportation basis and the structure of simulations and secondly the basic definition of instruments Araponga and Sabiá; Section 3 contains the results of the proposed simulations; In Section 4, we present paper conclusions.

## 2. Simulations and Instruments

In this present Section, we briefly describe neutron transportation and show proposed simulations to compare the neutron transportation efficiency by using a single (transmitted flux) or a split guide (split flux) at a hypothetical instrument position at the Neutron Guide Hall. Such scenarios of neutron transportation are investigated by considering different guide geometries and also symmetrical and asymmetrical coating inside curved guides. After this, we present a proposal of design for two priority instruments from the suite of 15 instruments of the neutron guide building of the RMB, namely Araponga and Sabiá, which consist of HRPND and SANS instruments, respectively. Such a proposal is a layout and technical description based on a study of state-of-art correspondent instruments.

### 2.1. Neutron Transportation Basis and Simulated Cases

It is given that a neutron beam flux after passing a beam port, which is a hole in the shielding, falls off according to a $r^{-2}$ behavior (where r is the source-instrument distance). This forces the allocation of instruments to be near to the reactor core, which consequently limits available space for operating them as much as makes it impossible to install a large quantity of equipment in the Neutron Hall of the facility. A widespread answer to overcoming this problem is neutron transportation based on optics instruments and geometry. Among these optic components, the supermirror is a fundamental device that allows neutron guides, which transport neutron beams to distances up to $100\ m$ away from the core without significant flux loss (some cases with 90% of transmission).

From the principles of geometric optics and quantum mechanics, supermirrors are developed to optimise neutron delivery in instruments at Neutron Guide Hall (NGH). Neutron scattering and interactions with surfaces like mirrors are ruled by solving the Schrödinger equation with a Fermi-pseudo-potential. In this context, a neutron guide just transports neutrons that have a maximum incident angle equal to the mirror critical angle. This angle depends on the same parameters $N$ and $b$ of Fermi-pseudo-potential, where $N$ is the atomic number density of material and $b$ the neutron scattering length. It is given by

$$sin\theta_c = \lambda\sqrt{\frac{Nb}{\pi}}, \qquad (3.1)$$

where $\theta_c$ is the critical angle and $\lambda$ is the wavelength of incident neutron. Here just neutrons with an incident angle with a value below $\theta_c$ are reflected by the mirror, otherwise, they are absorbed or transmitted.

In these terms, the relation between critical angle and wavelength varies according to the mirror element, which has correspondent values of $N$ and $b$. The neutron scattering length density is obtained for a natural element that corresponds to an element itself and its isotopes. By considering this, Nickel is the element of the periodic table with the highest $b$ value, which makes it the most appropriate material for building a mirror of neutrons. This scenario can be improved using a system of layers that enhance critical angle and consequently the transmitted flux through neutron guide. Besides, once a critical angle is small the approximation $sin\theta_c \approx \theta_c$ is also considered. The critical angle (in radians) of a supermirror is

$$\theta_c = 1.73 \times 10^{-3} m\lambda, \qquad (3.2)$$

where $m$ is the index used to classify supermirror according to its reflectivity. That is, $m$ is the ratio between the critical reflection angles of the supermirror and a Nickel coating surface. From the previous equation, we observe that is possible to balance transmitted flux wavelength shape according to used supermirrors. The linear dependence of critical angle and neutron wavelength dictates that colder neutrons are more easily transmitted than thermal ones. From this, we verify a different wavelength profile in the neutron guide entrance and exit. However, flux divergence and guide illumination are essential to shaping the final wavelength profile as much as neutron wavelength. This happens because large divergence values indicate that many neutrons of original flux do not hit the guide side with an incident angle less or equal to the critical angle.

When one considers a guide system entrance fully illuminated, it is possible to guarantee that a solid angle of $4\theta_c^2$ is, following the Liouville theorem, transported through such a system. Then the combination of accepted divergence and the $m$ index plays a fundamental role in the final wavelength profile and flux intensity at instrument positions. Curved guides, which are mainly used to avoid gamma rays and epithermal neutrons by excluding direct line-of-sight, also play an important role in the accepted flux since each neutron wavelength possesses a different efficiency of transmission [8,9].

According to the Acceptance Diagram (AD) approach, it is possible to model a guide that excludes line-of-sight utilizing guide curvature, width, and surface coating ($m$) next to a specific neutron wavelength. This is guaranteed by imposing that curved guide length is longer than characteristic length, which is given by

$$L_c = \sqrt{8W\rho}, \qquad (3.3)$$

where $W$ is guide width, and $\rho$ is guide curvature. Besides, the guide index coating $m$ is taken into account in guide neutron transport from the relation between neutron wavelength and characteristic wavelength, which, at least for Nickel coating, is described as

$$\lambda_c = \frac{1}{1.73 \times 10^{-3} m}\sqrt{\frac{2W}{\rho}}. \qquad (3.4)$$

In short, when $\lambda < \lambda_c$, neutrons are transported less efficiently (less than 66%) by considering just garland reflection regime and, on the other hand, they are transported more efficiently (more than 66%) for $\lambda > \lambda_c$, when both zig-zag and garland regimes are possible. In this sense, guides should be designed to guarantee that wanted instrument neutrons possess longer wavelengths than the characteristic wavelength. Inside such formalism, there is also a branch that considers the neutron transportation by curved guides without identical outer (concave) and inner (convex) surface coatings. Differently from straight guide AD, efficiency transportation areas (filling factor) are ruled by parabolical equations in the position-divergence space phase [8,9]. These equations dictate that grazing angles of the outer surface are always larger than their correspondent of the inner surface (considering a given trajectory). In this way, there is a limit where the inside surface coating index ($m_{in}$) can be less than the outer coating index ($m_{out}$) without losing transport efficiency. The upper limit of this range is written by a finite wavelength $\lambda'$ [10,11]. It is given by

$$\lambda' = \frac{m_{out}}{\sqrt{m_{out}^2 - m_{in}^2}} \lambda_c, \qquad (3.5)$$

where $\lambda_c$ comes from the Equation 3.4, but with $m = m_{out}$. From the Equation 3.5, we observe that $\lambda' \to \infty$ when $m_{in} \to m_{out}$ and $\lambda' \to \lambda_c$ for $m_{out} \gg m_{in}$. On the other hand, it is not valid for $\lambda < \lambda_c$, which regime is ruled just by garland reflection. In other words, when neutron wavelengths follow the relation $\lambda > \lambda_c$, it is ensured that their transport is more efficient than 66% and that only index $m_{out}$ plays a role in neutron transportation efficiency [10,11]. In this way and from an engineering point of view, it is possible to deliver the same range of neutron flux profile, but with fewer costs. Nevertheless, these properties are bound to a previous system definition. Since the guide width is much smaller than its length, we observe that supermirror of top and bottom of the guide does not interfere with those relations. For practical purposes, simulations in this paper are carried out considering both surface coatings (of top and bottom) with $m = 2$.

The definition of a guide system is a difficult task and firstly depends on previous sketched instruments of the project, which makes any study of guide system construction or improvement unique and particular. As a consequence, some fundamental and global aspects are frequently not present in the literature. In this sense and since OPAL is a reference for the RMB project, simulations of a generic guide system based on OPAL configuration are carried out to evaluate the difference between the transmitted and split flux to be delivered about $15\ m$ from the Bunker exit. In other words, we investigate two different scenarios based on two reference facilities, namely OPAL and FRM-II. From the former, we explore the configuration where the use of upstream flux that comes, for example, from Wombat is employed to deliver neutrons to Echidna. On the other hand, from the latter facility, we study the performance of guides based on its concept of split guides to distribute neutron flux for all instruments.

In simulations and this paper, we adopted the same nomenclature of OPAL guides. The portion of the simulated guides inside the Bunker, and sketched in Figure 2, is the same for simulations of split and transmitted flux. To save space between guides at the NGH, the central guide TG2 is, at least in this approach, straight and on the other hand, the edge guides TG1 and TG3 are curved in opposite directions, i.e., TG1 is bent to the right side and TG3 to the left side according to Figure 2 reference notation. The part of the guides inside the Bunker is simulated with a $40\ m$ long and section area of $30 \times 5\ cm^2$. Besides, curved guides possess a curvature of $4.5\ km$, which practically excludes the direct line-of-sight.

In cases with Bunker curved guides, their length and curvature guarantee no direct line-of-sight themselves. For other cases with curved guides shorter than the characteristic length (Equation 3.3), the line-of-sight is avoided through the composition of long straight guides and curved ones [12]. The rest of the guides continues through NGH and their geometry and characteristics depend on each simulated case.

In the simulations of split guides, we propose two types of equal division, a vertical and a horizontal one. These divided guides, which respectively possess sections areas of $30 \times 2.5\ cm^2$ and $15 \times 5\ cm^2$, are $15\ m$ long to feed neutron beams to a hypothetical place of an instrument at the NGH. All simulated cases configuration are present in Table 1, where the second column specifies the type of split, the third the shape of the primary guide (inside the Bunker), the forth the shape of the secondary guide (inside NGH), the fifth the connection between both parts where neutrons pass and finally the last column shows a chosen label for each case, which are normal capitals letters (A, B, C…). Here, we refer to guide shape as being three possible options, i.e., left-side curved, straight and right-side curved with a curvature of $4.5\ km$ for both curved options.

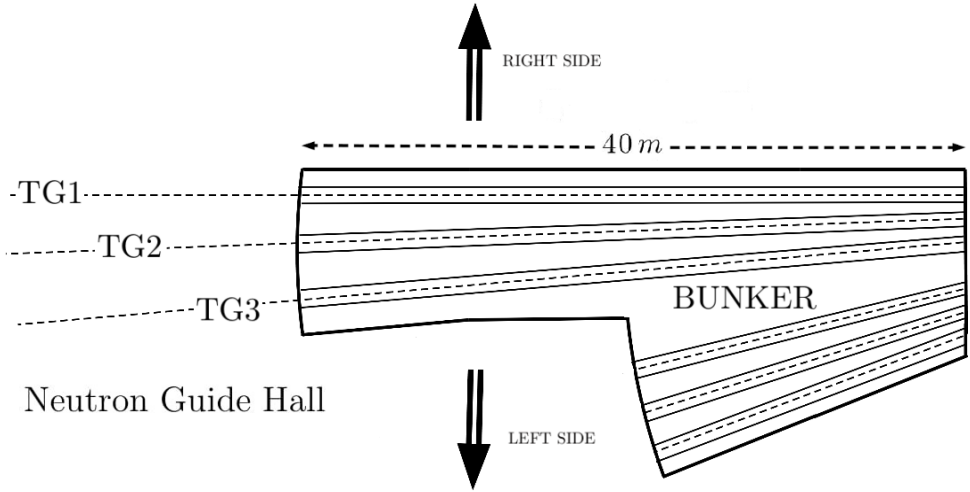

**Figure 2.** Top-view sketch of the Bunker and the Neutron Guide Hall of RMB. Primary guides are inside Bunker and possess $40\ m$ of length, where TG1 is a right-side curved guide, the TG2 is a straight guide and TG3 is a left-side curved guide. Both curved guides, TG1 and TG3, have curvature of $4.5\ km$.

The other simulation that we have carried out corresponds to the analysis of transmitted flux. These cases have the same initial guides inside the Bunker, but a different arrangement for guides allocated at the NGH. These guide parts inside the Bunker and NGH, like in the split guide simulations, are also referred to as primary and secondary guides, respectively. The secondary guides of this simulation are $6\ m$ long and can be curved (as in previous cases, to left or right side and with $4.5\ km$ of curvature) or straight. Secondary guides are followed by a collimator (of divergence $10'$ or open) and a pyrolytic graphite HOPG (002) monochromator. To obtain the same distance of $15\ m$ away from the Bunker, just like in split guide simulation, a straight guide of $8\ m$ is positioned just after the monochromator device (collimator and distances to monochromator have about $1\ m$).

Table 2 contains all simulation cases of transmitted flux. It is organized in a similar way of Table 1, but the second column contains primary collimator divergence, following the classical diffractometer arrangement of Caglioti [13]. Besides, there is no column describing the secondary guide connection since guides are not split in these simulations, i.e., NGH guides still have a $30 \times 5\ cm^2$ section area like Bunker guides. Here, all labels are distinguished from previous cases by capital letters with prime symbol (A', B', C'...).

There are two different sketches shown in Figures 3 and 4 that are representing those cases contained in Tables 1 and 2, respectively. They represent the sequel of possible components used in each simulation case. Figure 3 exhibits a layout of split flux cases, where the first frame contains all three options of guides inside the Bunker. Two possible frames represent sequentially the split type of each case, namely the vertical and horizontal split. The other two frames consist of guides option at the NGH, where their dimensions depending on split type are different. In Figure 4, which represents transmitted flux cases, the first frame shows Bunker guides possibilities as previously described in Figure 3. The second frame also contains the other three options of secondary guides, allocated at NGH. In the sequence, two frames represent possible collimation divergence, i.e., $10'$ and an open collimator. They are followed by the final part of the component until instrument place at the NGH, namely the monochromator and a straight guide.

**Table 1**: Configurations of split flux simulation cases. They are classified according to flux division type, horizontal and vertical, and to primary (inside Bunker) and secondary (inside NGH) guides side curvature. Each case corresponds to normal capital letters (A, B, C…) for thermal sources simulations.

| Simulation | Flux Split Type | Primary Guide (40 m) | Secondary Guide (15 m) | Secondary Guide Connection | Case |
|---|---|---|---|---|---|
| SPLIT | Horizontal | TG2 - Central (Straight) | Left (Curved to Left Side) | 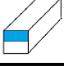 | A |
| | | | Right (Curved to Right Side) | 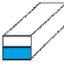 | B |
| | | TG3 - Left (Curved to Left Side) | Left (Curved to Left Side) | 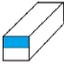 | C |
| | | | Straight (no curvature) | 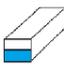 | D |
| | | | Right (Curved to Right Side) | 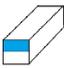 | E |
| | | TG1 - Right (Curved to Right Side) | Left (Curved to Left Side) | 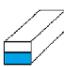 | F |
| | | | Straight (no curvature) | 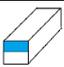 | G |
| | | | Right (Curved to Right Side) | 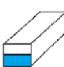 | H |
| | Vertical | TG2 - Central (Straight) | Left (Curved to Left Side) | 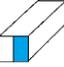 | I |
| | | | Right (Curved to Right Side) | 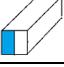 | J |
| | | TG3 - Left (Curved to Left Side) | Left (Curved to Left Side) | 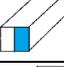 | K |
| | | | Straight (no curvature) | 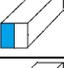 | L |
| | | | Right (Curved to Right Side) | 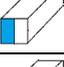 | M |
| | | TG1 - Right (Curved to Right Side) | Left (Curved to Left Side) | 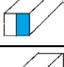 | N |
| | | | Straight (no curvature) | 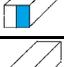 | O |
| | | | Right (Curved to Right Side) | 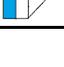 | P |

Figures 5 and 6 are examples of cases E and E', respectively. Here, Figure 5 is a sequence of a primary guide curved to the left side with a horizontal split and a guide curved to the right side, connected at its upper half (crosshatched area). On the other hand, Figure 6 is formed by the same primary and secondary guides of the previous case but followed by a 10′ collimator, a pyrolytic graphite HOPG (002) monochromator, and then a straight guide.

**Table 2**: Configurations of transmitted flux simulation cases. They are classified according to primary collimator divergence (horizontal), 10′ and open (4$^o$ 5′), and to primary and secondary guides side curvature. Following the previous nomenclature logic, cases of transmitted flux are described with capital letters with prime symbol (A', B', C'...).

| Simulation | Primary Collimator | Primary Guide (40 m) | Secondary Guide (6 m) | Case |
|---|---|---|---|---|
| Transmission | 10′ | TG2 - Central (Straight) | Left (Curved to Left Side) | A' |
| | | | Right (Curved to Right Side) | B' |
| | | TG3 - Left (Curved to Left Side) | Left (Curved to Left Side) | C' |
| | | | Straight (no curvature) | D' |
| | | | Right (Curved to Right Side) | E' |
| | | TG1 - Right (Curved to Right Side) | Left (Curved to Left Side) | F' |
| | | | Straight (no curvature) | G' |
| | | | Right (Curved to Right Side) | H' |
| | Open (4$^o$ 5′) | TG2 - Central (Straight) | Left (Curved to Left Side) | I' |
| | | | Right (Curved to Right Side) | J' |
| | | TG3 - Left (Curved to Left Side) | Left (Curved to Left Side) | K' |
| | | | Straight (no curvature) | L' |
| | | | Right (Curved to Right Side) | M' |
| | | TG1 - Right (Curved to Right Side) | Left (Curved to Left Side) | N' |
| | | | Straight (no curvature) | O' |
| | | | Right (Curved to Right Side) | P' |

Since we are interested in comparing neutron beam intensities at instrument position for split and transmitted fluxes, simulations have been carried out for all cases of Tables 1 and 2 for five different thermal sources. Inside McStas software, it is possible to mimic some known reactor sources by following the temperature and intensity parameter that describes a Maxwellian distribution. For these virtual sources, we have used parameters present in the McStas website to

– 9 –

describe the thermal sources of FRM-II, HZB, ILL, and a Triga reactor model [14]. Also, we used an MCNP output file to simulate the thermal source of the Brazilian IEA-R1 reactor.

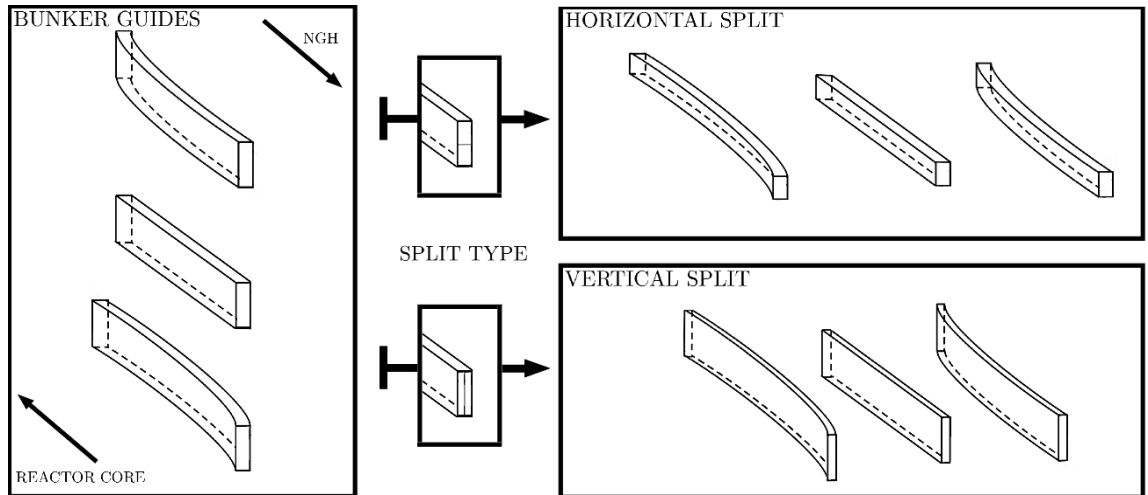

**Figure 3.** A sketch of all possible settings of split flux simulations shown in Table 1. The first frame contains the three guide possibilities of the Bunker, i.e., left-curved, straight and right-curved guides. It is followed by two types of split and then by the corresponding set of guide options at the NGH (last frame), which are guides that possess the same options present in the first frame but with different dimensions. Note that there is no combination of two straight guides to avoid direct line-of-sight.

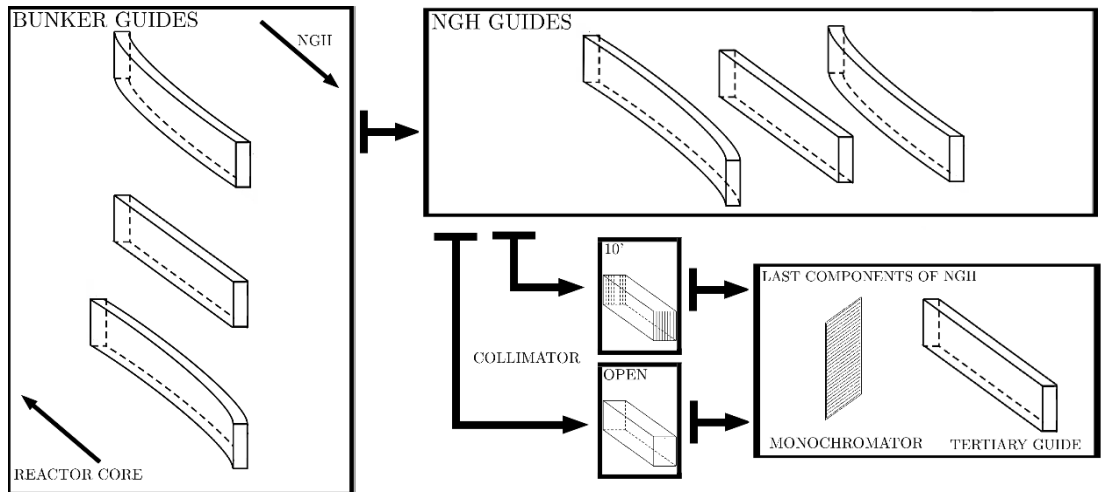

**Figure 4.** A sketch of all possible settings of transmitted flux simulations shown in Table 2. The first frame contains the three guide possibilities of the Bunker, i.e., left-curved, straight and right-curved guides. It is followed by a frame containing possible options of secondary guides, which also can be left or right-side curved, or straight. The next step shows two types of collimator that can have a divergence of $10'$ or be an open one. Both options are succeeded by a unique set formed by a monochromator and a straight guide. Note that there is no combination of two straight guides to avoid direct line-of-sight.

Through simulation results, we are able, as it will be shown in Section 3, to infer the independence of the relative efficiency of a split and transmitted flux on the source wavelength profile. In these terms, such a result can be used in future simulations for the RMB system guide definition, since by the time there is only a correspondent TG2 MCNP output file (provided by INVAP) available. Particularly, it makes possible to deduce the behavior of other thermal and cold guides, i.e., TG1 and TG3, and also CG1, CG2, and CG3, respectively.

We use these configurations in two different types of simulations, which is divided into two steps, where the first one consists of running all simulations out by using all inside surface coating indexes with the same value, namely $m = 2$. Once the relation between efficiency cases are independent on profiles and intensities of the source, and geometry (source size and distance to guide system entrance), we select configurations A, C, D, E, I, K, L, and M (and their lined correspondent transmitted cases) for simulations using the ILL source and with different $m_{out}$ and $m_{in}$.

This second set of simulations are performed for $m_{out} = 2.5$ and $m_{in} = 2$ and also $m_{out} = 2$ and $m_{in} = 1.5$. For comparation purpose, we present these results with the previous simulation cases of ILL source with $m_{out} = 2$ and $m_{in} = 2$. As in previous simulations, all top and bottom surfaces of curved guides have $m = 2$. All straight guides of the first simulation, as much as the second one, possess fixed surface indexes of $m = 2$.

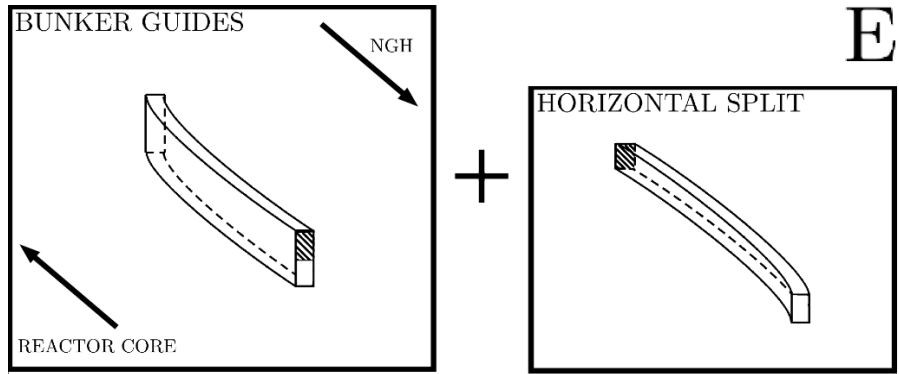

**Figure 5.** A sketch of case E shown in Table 1. In the first frame, there is a primary guide split horizontally into two halves. It is curved to the left side and is $40\ m$ long. The second frame contains the secondary guide, which is connected to the upper part of the primary guide. This guide is curved to the right side and has a length of $15\ m$. Primary and secondary guides possess section areas of $30 \times 5\ cm^2$ and $15 \times 5\ cm^2$, respectively.

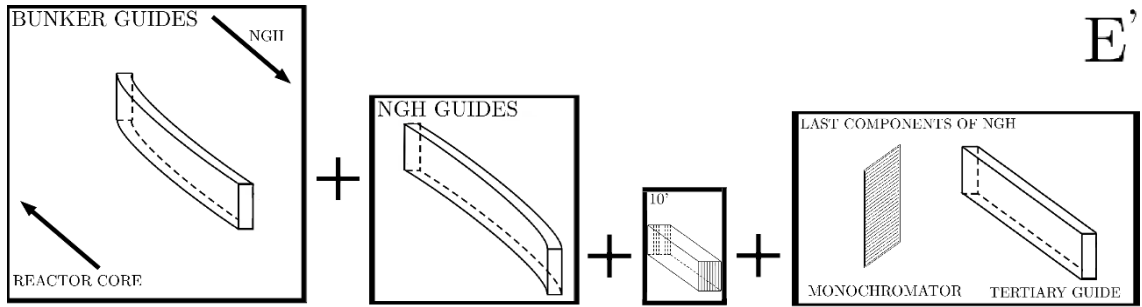

**Figure 6.** A sketch of case E' shown in Table 2. The first frame contains the primary guide allocated inside the Bunker. It is curved to the left side and $40\ m$ long. The second frame consists of a secondary guide that is curved to the right side and is $6\ m$ long. The third frame represents a $10'$ collimator. The forth, and last frame, is composed of a pyrolytic graphite HOPG (002) monochromator and an $8\ m$ straight guide. Guides of all stages possess section areas of $30 \times 5\ cm^2$.

### 2.2. Instruments

In this Subsection, we present the basic configurations of the Araponga and Sabiá instruments, two of the suite's main instruments to be housed in the NGH, which is part of the N02. This building is a modern facility with auxiliary laboratories for sample preparation and analysis. In the basic design, N02 will house a suite of 15 instruments, named with birds of the Brazilian fauna. The instruments are Neinei (Neutrography and Tomography), Lenheiro (Laue Diffractometer), Rendeira (Reflectometer), Tucano (Tomography), Tororó (Cold 3-Axis Spectrometer), Tico-Tico (Time-of-flight Reflectometer), Saracura (Very Small-Angle Neutron Scattering), Beija-Flor (Backscattering Diffractometer), Dançador (Diffuse Scattering Diffractometer), Sabiá (Small-Angle Neutron Scattering), Flautin (High Intensity Diffractometer), Araponga (High Resolution Diffractometer), Estrelinha (Thermal 3-Axis Spectrometer), Siriema (Residual Stress Diffractometer) and Mutum (Single Crystal Diffractometer). In this paper, we present the basic configurations of the Araponga and Sabiá instruments.

### 2.2.1. Araponga-HRPND

Araponga-HRPND is an instrument based mainly on correspondent High-resolution diffractometer Echidna at OPAL [15]. Aspects of HZB diffractometers E6 and E9 as vertical and horizontal focusing monochromators are also considered. The standard instrument (shown in Figure 7) is sequentially composed of main guide, primary collimator, vertical focusing monochromator, secondary guide, secondary collimator, sample position (sample environment), radial collimator and detector. The state-of-art of a high-resolution diffractometer dictates that such an instrument should be located in a thermal guide, where wavelength peak stays between 1 Å and 2 Å. Due to forward neutron scattering preference behavior and monochromators d-spacing, we have that take-off angles have mostly high values, e.g., between 90° and 140°.

**Table 3**: Technical parameters of HRPND - Araponga.

| Position | Thermal Guide (TG1) |
| --- | --- |
| Collimation | $5' \leq \alpha_1 \leq 18'$, $10' \leq \alpha_2 \leq 20'$, $\alpha_3 = 5'$ |
| Wavelength range (peak) | $1\text{ Å} < \lambda < 3\text{ Å}$ ($\lambda_{peak} = 1.5\text{ Å}$) |
| Take-off angle | $90^o \leq 2\theta \leq 140^o$ |
| Resolution Δd/d | $\sim 10^{-4}$ (0.1%) |
| Neutron Flux (at sample position) | $\sim 10^7 \, n/cm^2 s$ |

According to the milestone work of Caglioti and collaborators [13] (and also Hewat application work [16]) on diffractometer assembly, one can estimate the full width at half maximum (FWHM) according to the divergence of the three collimators combined with the monochromator mosaicity. This estimation allows optimising a high-resolution diffractometer configuration to deliver a minimum necessary flux at the sample combined with a fine angular resolution to better distinguish diffracted peaks at the final detector. Following these principles and the state-of-art diffractometer Echidna, SPODI, E9, and D2B, we predefine primary collimator between $5'$ and $18'$, secondary collimator between $10'$ and $20'$ and tertiary to be equal to $5'$. We also assume a monochromator mosaicity around $33'$, as that on Echidna instrument [15].

A maximum flux of $10^7\ n/cm^2 s$ at the sample place of the Araponga is expected from reference diffractometers Echidna and D2B as much as a fine d-spacing resolution of 0.1%. Here, all these predefined values of the Araponga diffractometer are presented in Table 3.

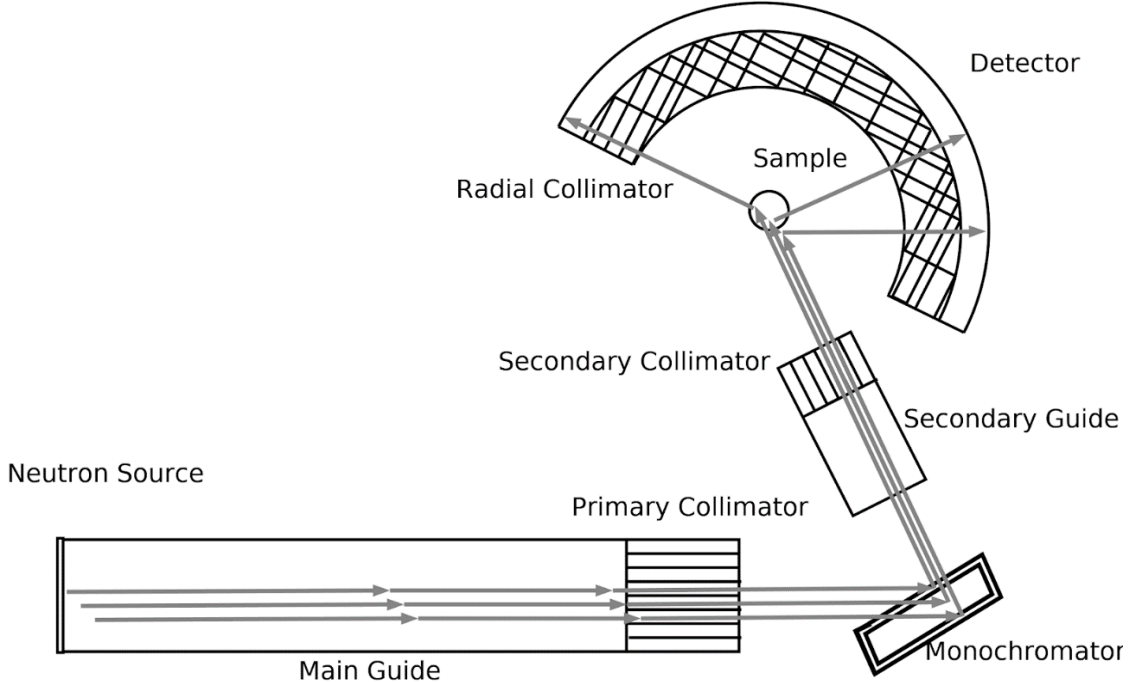

**Figure 7.** Schematic view of Diffractometer-Araponga.

### 2.2.2. SANS-Sabiá

SANS-Sábia is based on the instruments Quokka and Bilby [17,18] at OPAL, which have similar components to the D33 instrument at ILL [19]. Figure 8 presents the basic layout of the instrument. The total length of SANS-Sabiá is $L_1 + L_2 = 40\ m$, where $L_1$ is the distance between the mechanical velocity selector and the sample, and $L_2$ the sample-detector distance. A summary of the main SANS-Sabiá parameters is found in Table 4. Following the design of most modern neutron scattering instruments, SANS-Sabiá has an estimated maximum flux of approximately $10^7\ n/cm^2 s$ for $\lambda = 4.5$ Å at the sample position. The aim is to achieve a high brightness beam in its basic design. Neutron transport is made by guides ($50\ mm \times 50\ mm$) coated by supermirrors with $m = 2$. The options of mechanical velocity selectors with resolutions $\Delta\lambda/\lambda = 6 - 20\ \%$ for $\lambda = 4.5$ Å are currently being evaluated. The polarization system will allow a detailed study of magnetic structures in transmission geometry, followed by a radio-frequency spin flipper. The collimation length is in the range $L_1 \leq 20$ m and the distance sample-detector can be $L_2 \leq 20$ m.

A $^3$He detector with $1\ m^2$ area will be used to detect neutrons scattered in the $2.5\ m$ diameter cylindrical tank. In the future, a set of sample environments will be developed to propose more complex studies.

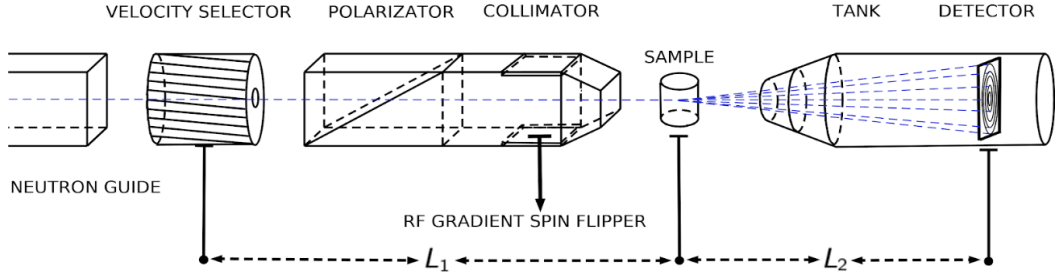

**Figure 8.** Schematic view of SANS-Sabiá.

**Table 4:** Technical parameters of SANS - Sabiá.

| Position | Cold Guide (CG1) |
| --- | --- |
| Wavelength | $4.5 \text{ Å} < \lambda < 30 \text{ Å}$ |
| Collimation length | $L_1 < 20 \, m$ |
| Polarization | $> 90 \%$ |
| Sample-detector distance | $L_2 < 20 \, m$ |
| Detector area | $1 \, m^2$ |
| Neutron Flux (at sample position) | $\sim 10^7 \, n/cm^2 s$ |

## 3. Results

Simulation cases of Tables 1 and 2 have been carried out and their results are organized as flux ratios and presented with bar charts in Figures 9 and 10 for simulations with $m_{out} = m_{in}$ and Figures 11 and 12 for $m_{out} > m_{in}$ as previously described. Ratios are obtained by dividing the flux at the exit of the guide system by the flux at the entrance of the primary guide. These ratios, which represent the efficiency of each case according to its correspondent source, allow us to compare all results independently of initial wavelength profile and reactor intensity. Since our first goal consists of comparing ratios of split and transmitted flux cases, values are also shown in Tables 5, 6, 7 and 8, respectively to Figures 9, 10, 11 and 12.

After analysing the results of proposed cases, we verify that guide system efficiency relation among different cases is independent of thermal source choices (Simulations utilizing cold neutron sources have also been carried out, and referred independence is maintained). From such results, we confirm the validity of the present study results on posterior simulations with the RMB source. From this scenario, we observe that the decreasing (increasing) sequence of orders, between the most and least efficient (least and most efficient), is always maintained regardless of the source used in the simulations.

By comparing both graphs of Figures 9 and 10, it is noticeable that the order of instruments in a transmitted flux guide is crucial in transporting neutrons efficiently to the last instrument. That is, the use of a $10'$ collimator, which corresponds, for example, to a setup of a high-resolution diffractometer, excludes the possibility of an upstream installation of a high-intensity instrument.

Notwithstanding, it is fundamental to guarantee that this flux at the ending of the guide system still contains neutron beams of the desired wavelength for the instrument, i.e., both monochromators have to be dissociated in wavelength selection in case of two diffractometers.

The vertical and horizontal splits are only equivalent when the incoming flux is homogeneous, i.e., when the primary guides are straight. Since straight guides do not exclude line-of-sight, we observe that configurations A, B, I, J, A', B', I' and J' are the most efficient from their correspondent case.

Cases, where primary guides are curved, impose less flux at the final position instrument at NGH than those with primary straight guides. Inside split cases, we observe an imbalance of neutron beam when the vertical split is applied. This happens because curved guides tend to accumulate more neutrons on the outer side according to garland reflections and to the AD approach. In this way, any guide configuration that selects the outer flux would be more efficient. This is noticeable by comparing cases L, M, N, and O with cases K and P. Since the acceptance of these curved guides do not depend on guide height, we observe that the horizontal split plays no role in the efficiency of both guide halves. Therefore, when cases with $10'$ divergence (A' - H') are not considered, just cases K and P are less efficient than transmitted ones.

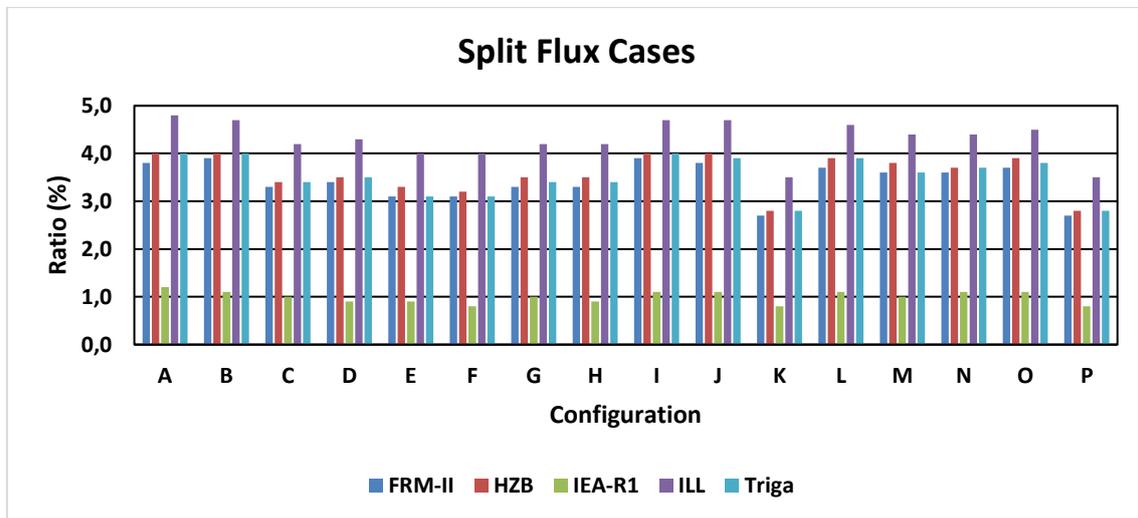

**Figure 9.** A bar chart of results of split flux from A to P cases. Numerical values are in Table 5. Y-axis possesses ratios (in percentage) between initial and final fluxes for each case and source. The X-axis shows labels of cases, where each one of them presents ratios of all five thermal sources used in simulations. Dark blue, red, green, purple and light blue colored bars represent FRM-II, HZB, IEA-R1, ILL, and Triga thermal sources, respectively.

By analysing transmitted cases, we observe that all results with an open collimator (I' - P') provide higher fluxes than their corresponding cases with collimation ($10'$). However, real instruments, like a high-intensity diffractometer, usually need some collimation before selecting wavelength. So, any non-open collimator would diminish flux at the end of the guide system and consequently would increase the difference between split and transmitted flux, turning the use of split cases more advantage, at least in terms of efficiency, than the transmitted ones.

As already expected, we observe that the use of collimators in a transmitted flux is crucial for efficiency at the last instrument position. In our simulation results, we found efficiencies of about three times less for cases of collimation of $10'$ than for cases of an open collimator. In these cases, the design of both instruments, i.e., monochromators, collimators, and other components, has necessarily to be correlated to guarantee fine measurements. Still considering this dependence, it is always necessary to allocate high-intensity instruments in the first position of the upstream flux at the guide system. However, cases have to be studied individually taking into account particular instrument characteristics and the corresponding costs.

For most comparative cases, we observe that split cases are more efficient than transmitted cases except those cases where vertical split selects the inner guide flux for instrument position at the NGH (K and P). In this scenario, there are two options to avoid a disbalance division of neutron beams and consequently keeping the same wavelength profile for both fluxes. The first consists of splitting guide horizontally since according to AD formalism there is symmetry on neutron distribution in the vertical axis (since guide curvature is in the horizontal plane). The second possibility is the use of a straight guide before splitting any part of the guide system. Nevertheless, a curved guide can be used to modify and filter specific wavelengths ranges depending on each instrument characteristic in different parts of NGH.

Results of the second part of simulations are presented in Tables 7 and 8, where values are plotted in Figures 11 and 12, respectively. By analysing such results, we first observe that the use of different coating indexes does not alter, at least for these configurations, the relation between efficiencies. In other words, the relation among all cases is the same in a way that the order of most to less efficient cases is sequentially maintained the same, e.g., case A, I, A' and I' are still the most efficient ones. This effect is similar to the independence of efficiency relations and the used source in simulated cases, where the sequence of most efficient geometries is fixed apart of the initial profile. In these terms, we conclude that we are able to infer the results of the first simulations as an available start approach for RMB guide system definition.

**Table 5**: Results of the split flux guide system for thermal neutron sources, namely FRM-II, HZB, IEA-R1, ILL and Triga Mark reactor. Values are also presented in Figure 9.

| Configuration | Split Flux Cases | | | | |
|---|---|---|---|---|---|
| | Thermal Source | | | | |
| | FRM-II | HZB | IEA-R1 | ILL | Triga |
| | Ratio (%) | Ratio (%) | Ratio (%) | Ratio (%) | Ratio (%) |
| A | 3.8 | 4.0 | 1.2 | 4.8 | 4.0 |
| B | 3.9 | 4.0 | 1.1 | 4.7 | 4.0 |
| C | 3.3 | 3.4 | 1.0 | 4.2 | 3.4 |
| D | 3.4 | 3.5 | 0.9 | 4.3 | 3.5 |
| E | 3.1 | 3.3 | 0.9 | 4.0 | 3.1 |
| F | 3.1 | 3.2 | 0.8 | 4.0 | 3.1 |
| G | 3.3 | 3.5 | 1.0 | 4.2 | 3.4 |
| H | 3.3 | 3.5 | 0.9 | 4.2 | 3.4 |
| I | 3.9 | 4.0 | 1.1 | 4.7 | 4.0 |
| J | 3.8 | 4.0 | 1.1 | 4.7 | 3.9 |
| K | 2.7 | 2.8 | 0.8 | 3.5 | 2.8 |
| L | 3.7 | 3.9 | 1.1 | 4.6 | 3.9 |
| M | 3.6 | 3.8 | 1.0 | 4.4 | 3.6 |
| N | 3.6 | 3.7 | 1.1 | 4.4 | 3.7 |
| O | 3.7 | 3.9 | 1.1 | 4.5 | 3.8 |
| P | 2.7 | 2.8 | 0.8 | 3.5 | 2.8 |

Considering the ratios of different coating configurations, we observe an expected behavior of efficiency increasing as much as $m_{out}$ or $m_{in}$ is raised. However, the point here is not to pick the most efficient set of supermirror indexes, since the most obvious choice would be to use the highest available $m$ value. Here, we investigate how to avoid using an improved value of $m_{in}$ in vain. In this study, we consider the entire neutron wavelength profile to infer configuration efficiency, but, since we are dealing with a thermal spectrum, these results are a reasonable first approach to compare guide properties.

According to Equation 3.5, we have that cases with $m_{out} = 2.5$ and $m_{in} = 2$, and $m_{out} = 2$ and $m_{in} = 1.5$ possess $\lambda'$ equal to 1.82Å and 2.06Å, respectively. Besides, by substituting $m_{out}$ values into Equation 3.4 it is possible to set $\lambda_c$ equal to 1.09Å and 1.36Å. These upper and lower limits dictate a specific wavelength range where neutrons are transported with the same efficiency of cases with $m_{out} = m_{in} = 2.5$ (1.09Å $\leq \lambda <$ 1.82Å) and $m_{out} = m_{in} = 2$ (1.36Å $\leq \lambda <$ 2.06Å), respectively. Both ranges of $\lambda'$ and $\lambda_c$ provide a more efficient neutron transport for the wavelength range that Araponga requires, i.e., 1Å $\leq \lambda <$ 3Å.

Since the ILL source profile contains the most of neutrons with wavelengths between 0.5 and 2.5, we observe just a slight difference between cases with $m_{out} = m_{in} = 2$ and $m_{out} = 2$ and $m_{in} = 1.5$. According to AD formalism, we expect an improved correspondence between cases with $m_{out} = 2$ and $m_{in} = 2$, and $m_{out} = m_{in} = 1.5$., when considering just range between $\lambda_c$ and $\lambda'$ instead of the entire wavelength profile. After considering that, it is natural to verify that the case with $m_{out} = 2.5$ and $m_{in} = 2$ is the most efficient. This occurs because of such configuration is equivalent to a curved guide with both curved surfaces coated with $m_{out} = m_{in} = 2.5$.

By checking Figures 11 and 12, we can also observe that the efficiency of cases A, I, A' and I' are not significantly improved for supermirrors with $m_{out} = 2.5$ and $m_{in} = 2$. This is because curved sections of these cases are at least six times shorter than curved parts of other configurations. From these results, we confirm that the use of different outer and inner supermirror indexes is fundamental for neutron transport optimisation since some scenarios are not improved for greater inner indexes. Here, split cases are still more efficient than transmitted ones as previous results already shown.

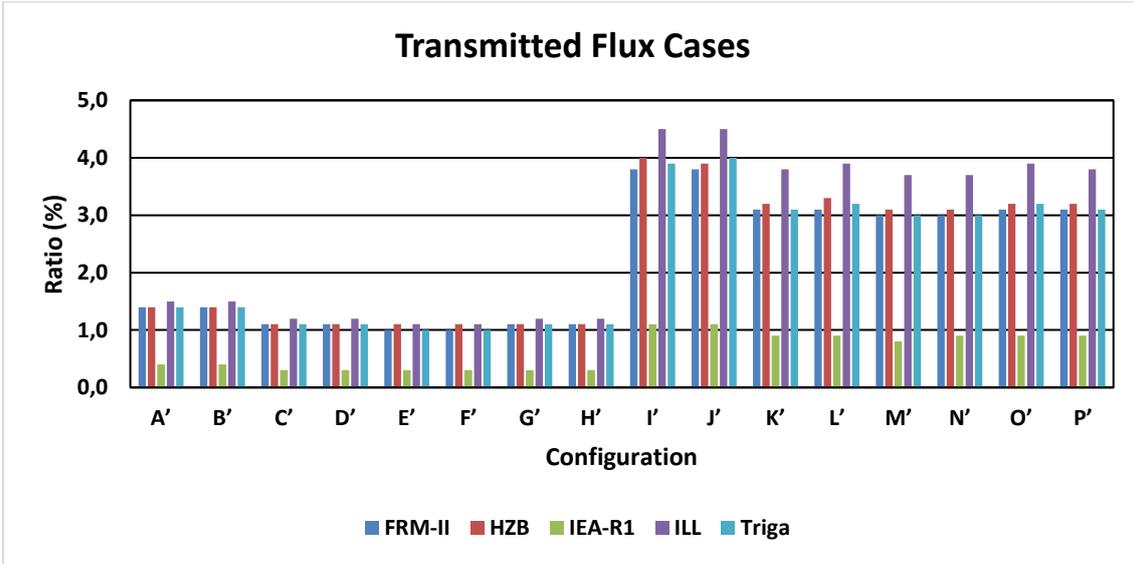

**Figure 10.** A bar chart of results of transmitted flux from A' to P' cases. Numerical values are shown in Table 6. The Y-axis possesses ratios (in percentage) between initial and final fluxes for each case and source. The X-axis shows labels of cases, where each one of them presents ratios of all five thermal sources used in simulations. Dark blue, red, green, purple and light blue colored bars represent FRM-II, HZB, IEA-R1, ILL, and Triga thermal sources, respectively.

There are still some aspects not addressed in proposed simulations and are crucial for guide system definition when considering engineering points. Naturally, configurations A, B, I, J, A', B', I' and J' provide higher fluxes to instruments since Bunker guides, which are the longest part of the system (40 $m$), are straight. The elimination of neutrons employing curved guides is a task that should be performed preferentially inside the Bunker, which has a proper shield coating to avoid gamma rays and epithermal neutrons to scape. Despite there is no line-of-sight for performed simulations, it is still desirable to circumvent the problem of excluding epithermal neutrons and gamma-rays in the NGH, otherwise, extra shielding would be necessary. Another possible way to solve this problem is through the use of filters. However, these solutions would bring additional concerns in terms of safety, costs and instrument properties, which demand further simulations on future works for checking the influence of the use of a filter on A, A', B, B', I, I', J and J' cases.

**Table 6**: Results of the transmitted flux guide system for thermal neutron sources, namely FRM-II, HZB, IEA-R1, ILL and Triga Mark reactor. Values are also presented in Figure 10.

| | Transmitted Flux Cases | | | | |
|---|---|---|---|---|---|
| | Thermal Source | | | | |
| Configuration | FRM-II | HZB | IEA-R1 | ILL | Triga |
| | Ratio (%) | Ratio (%) | Ratio (%) | Ratio (%) | Ratio (%) |
| A' | 1.4 | 1.4 | 0.4 | 1.5 | 1.4 |
| B' | 1.4 | 1.4 | 0.4 | 1.5 | 1.4 |
| C' | 1.1 | 1.1 | 0.3 | 1.2 | 1.1 |
| D' | 1.1 | 1.1 | 0.3 | 1.2 | 1.1 |
| E' | 1.0 | 1.1 | 0.3 | 1.1 | 1.0 |
| F' | 1.0 | 1.1 | 0.3 | 1.1 | 1.0 |
| G' | 1.1 | 1.1 | 0.3 | 1.2 | 1.1 |
| H' | 1.1 | 1.1 | 0.3 | 1.2 | 1.1 |
| I' | 3.8 | 4.0 | 1.1 | 4.5 | 3.9 |
| J' | 3.8 | 3.9 | 1.1 | 4.5 | 4.0 |
| K' | 3.1 | 3.2 | 0.9 | 3.8 | 3.1 |
| L' | 3.1 | 3.3 | 0.9 | 3.9 | 3.2 |
| M' | 3.0 | 3.1 | 0.8 | 3.7 | 3.0 |
| N' | 3.0 | 3.1 | 0.9 | 3.7 | 3.0 |
| O' | 3.1 | 3.2 | 0.9 | 3.9 | 3.2 |
| P' | 3.1 | 3.2 | 0.9 | 3.8 | 3.1 |

**Table 7:** Results of the split guide system for the thermal source ILL with different curved surface coatings, namely, $m_{out} = 2.5$ and $m_{in} = 2$, $m_{out} = 2$ and $m_{in} = 2$, and $m_{out} = 2$ and $m_{in} = 1.5$.

| | ILL Split Cases | | |
|---|---|---|---|
| Configuration | $m_{out} = 2.5/m_{in} = 2$ | $m_{out} = 2/m_{in} = 2$ | $m_{out} = 2/m_{in} = 1.5$ |
| | Ratio (%) | Ratio (%) | Ratio (%) |
| A | 5.0 | 4.8 | 4.7 |
| C | 4.8 | 4.2 | 4.2 |
| D | 4.8 | 4.3 | 4.2 |
| E | 4.7 | 4.0 | 3.9 |
| I | 4.9 | 4.7 | 4.6 |
| K | 4.2 | 3.6 | 3.5 |
| L | 4.9 | 4.6 | 4.6 |
| M | 4.9 | 4.5 | 4.3 |

Taking these previous cases apart, the most promising scenarios to be considered in RMB guide definition are cases C, D, G, H, K, L, O, and P for split configurations and C', D', G', H', K', L', O', and P' for transmitted ones. According to literature, the asymmetry imposed on neutrons by curved guides can be a difficult issue on instrument design. Generally, this problem is circumvented by the use of asymmetric convex and concave surface coating and/or by inserting a straight guide after curved one.

**Table 8:** Results of the transmitted guide system for the thermal source ILL with different curved surface coatings, namely, $m_{out} = 2.5$ and $m_{in} = 2$, $m_{out} = 2$ and $m_{in} = 2$, and $m_{out} = 2$ and $m_{in} = 1.5$.

| Configuration | ILL Transmitted Cases | | |
|---|---|---|---|
| | $m_{out} = 2.5/m_{in} = 2$ | $m_{out} = 2/m_{in} = 2$ | $m_{out} = 2/m_{in} = 1.5$ |
| | Ratio (%) | Ratio (%) | Ratio (%) |
| A' | 1.5 | 1.5 | 1.5 |
| C' | 1.3 | 1.2 | 1.2 |
| D' | 1.3 | 1.2 | 1.2 |
| E' | 1.3 | 1.1 | 1.1 |
| I' | 4.5 | 4.5 | 4.5 |
| K' | 4.3 | 3.8 | 3.7 |
| L' | 4.3 | 3.9 | 3.8 |
| M' | 4.2 | 3.7 | 3.7 |

Considering this problem, we highlight those configurations that possess straight secondary guides, which are slightly more efficient and also improve neutron divergence and distribution, namely configurations D, D', G, G', L, L', O and O'. From them, transmission cases are quite similar to OPAL cold and thermal guide configurations, i.e., D', G', L' and O'. In this sense, we should analyse in future works if there are motivations, which are out of simulations, that could unviable the use of split guides to feed neutrons to NGH instruments. Considering that cases with $m_{in} = 1.5$ are equal or just 0.1% less than those with $m_{in} = 2$ for both transmitted and split scenarios, we should also perform these sets of simulations with different supermirror indexes (up to $m = 3$) to optimise neutron transportation.

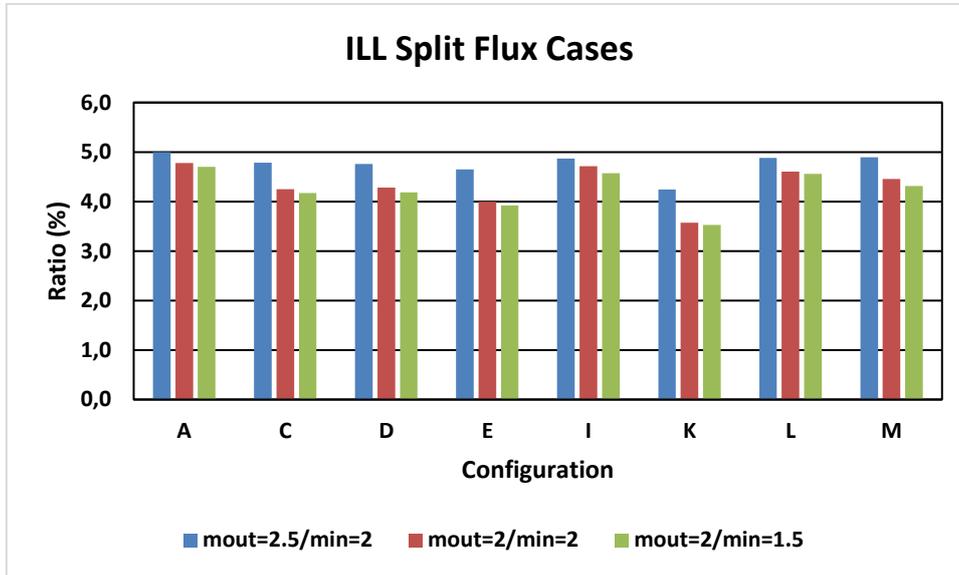

**Figure 11**. A bar chart of results of split flux from A, C, D, E, I, K, L and M cases for ILL thermal source and with different curved surface coatings. The X-axis shows labels of cases, where each one of them presents ratios of all three guide supermirror sets. Blue, orange and grey colored bars represent cases with $m_{out} = 2.5$ and $m_{in} = 2$, $m_{out} = 2$ and $m_{in} = 2$, and $m_{out} = 2$ and $m_{in} = 1.5$, respectively.

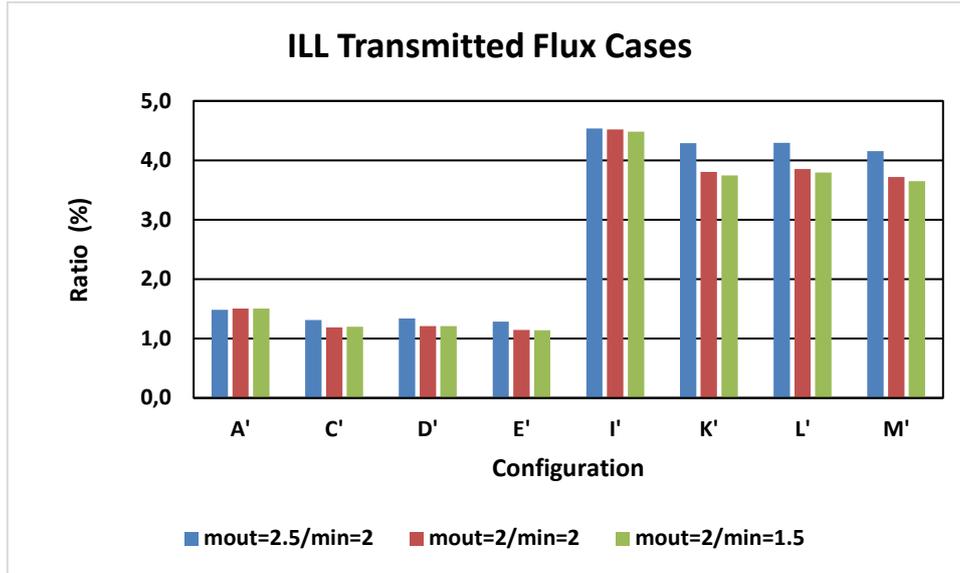

**Figure 12**. A bar chart of results of transmitted flux from A', C', D', E', I', K', L' and M' cases for ILL thermal source and with different curved surface coatings. The X-axis shows labels of cases, where each one of them presents ratios of all three guide supermirror sets. Blue, orange and grey colored bars represent cases with $m_{out} = 2.5$ and $m_{in} = 2$, $m_{out} = 2$ and $m_{in} = 2$, and $m_{out} = 2$ and $m_{in} = 1.5$, respectively.

## 4. Conclusions

In this study, we investigate some basic aspects of the use of split and single guides to distribute neutron beams to two instruments. This investigation consists of simulating a set of simplified guide systems employing ray-tracing Monte Carlo with the McStas software. A prime set of simulations is carried out for five different sources considering curved guides with equal concave and convex coating surfaces ($m_{out} = m_{in} = 2$). Additional simulations are performed taking into account asymmetric scenarios, where outer surface sides have higher coating indexes than inner ones ($m_{out} > m_{in}$). From these first results, we intend to perform more realistic simulations employing RMB proper source with an MCNP input file, which is provided by INVAP, and higher coating indexes, e.g., $m = 3$. We guarantee such employment from the independence of efficiency behavior among proposed cases next to different sources. That is, the scenario of the relative efficiency of cases is fixed for any choice of the source, e.g., the most efficient configuration is always the same, independently of the source. In this way, the most appropriate geometry cases obtained in this study are also the most promising approaches for the RMB guide definition.

From the presented simulation results, we conclude that the split option is the most efficient to transport neutron beams to instrument final position. However, considering that the efficiency of transmitted and split cases differs about 20% for some configurations, it is highly necessary to take the cost into account in defining the guide system. This occurs because split guides require more coating than single ones, which consequently make them more expensive. Here, additional discounts on total costs can also be inferred from the results of different indexes curved guide cases that indicate equivalent efficiencies for both symmetric and asymmetric surface coating.

Straight Bunker guide cases are more efficient, but from an engineering point of view, it is necessary to avoid excluding neutrons out of Bunker shielding. This requires additional investigations including filters inside the neutron transportation system. Considering cases with curved Bunker guides, we verify that cases with straight secondary guides are the most efficient for all configuration scenarios (both split types and both collimator divergences). When secondary guides are curved in the same direction as the primary one, we see that vertical splitting is an unfavorable scenario, since neutrons inertially tend to stay at the outer part of the guide. For horizontal split and transmitted cases, the use of primary and secondary guides bent on the same side is more efficient than vertical split cases. Taking the distribution asymmetry imposed by bent guides into consideration, there is no significant gain in efficiency of curved on straight secondary curved cases that justify going through these problems.

Despite an apparent disadvantage of transmitted flux next to split flux, it is worth to analyse the instrument method of selecting monochromatic neutrons. A fundamental difference is noticeable in instruments as Araponga and Sabiá, where the former is designed to use a monochromator and the latter a velocity selector. With the monochromator, the selected neutron beam is redirected, letting the rest of the neutrons to go upstream. On the other hand, the velocity selector eliminates all neutrons with a different selected wavelength, letting just selected ones. From this point of view, the process of utilizing transmitted flux avoids wasting neutrons not used in the first instrument, but at the cost of both instrument set configuration being bound.

Last but not least, we conclude that configuration D, D', G, G', L, L', O and O' are the most promising scenarios for RMB guides definition to be performed soon. Configurations L' and O' are quite similar to OPAL TG1 guide geometry and this fact opens a reasonable preceding for studying and applying these cases on RMB guides. Nevertheless, it is still necessary to compare this geometry with split guide configurations since we observe that their cases are mostly more efficient than transmitted ones according to each case. By taking engineering aspects into account on future studies, we need to check which conditions are fundamental and could avoid the use of split guides cases besides guide system definition based on configurations L' and O'. Otherwise, the OPAL set of neutron guides and configurations would be the most appropriate and prudent starting point for future simulations considering the RMB source input file. These complete works will mandatorily take into account the determination of guide curvature and width, instrument position, the use of filters, the presence of line-of-sight, problems with shielding, etc.

## Acknowledgments

The authors are thankful to the technical coordinator of the RMB project, Dr. J.A. Perrotta. APSS and LPO also would like to thank CNPq for financial support under grant numbers 381565/2018-1 and 380183/2019-6, respectively.

## References

[1] Perrotta, J. A., and Soares A. J. RMB: The New Brazilian Multipurpose Research Reactor. *ATW. International Journal for Nuclear Power* **60** (2015) 30-34.

[2] Villarino, E., & Doval, A. (2011). INVAP's Research Reactor Designs. *Science and Technology of Nuclear Installations (Online)*, **2011** (2011), 6.

[3] https://www.lnls.cnpem.br/activities/user-statistics/ (accessed 03 Feb. 20).

[4] Perrotta, J. A., Soares, A. J., Genezini, F. A., Souza, F. A., Franco, M. K. K. & Granado, E. Future Perspectives for Neutron Beam Utilization in Brazil. *Neutron News*. **25:4** (2014) 3-5.

[5] Steichele, E. Experimental installations and instruments at the FRM-II. in *Proceedings of the 7. meeting of the International Group on Research Reactors - International Atomic Energy Agency (IAEA)*, (1999).

[6] Lefmann, K. and Nielsen, K. A general software package for neutron ray-tracing simulations. *Neutron News*. **10:3** (1999) 20-23.

[7] de Oliveira, L.P. *et al*. Monte Carlo simulations of the S-shaped neutron guide. *JINST*. **15** (2020) P01012.

[8] Mildner, D. Acceptance diagrams for curved neutron guides, *Nucl. Instrum. Meth.* A **290** (1990) 189.

[9] Maier-Leibnitz, H. and Springer, T. The use of neutron optical devices on beam-hole experiments. J. Nucl. Ener. A/B **17** (1963) 217.

[10] Cook, J.C. Design and estimated performance of a new neutron guide system for the NCNR expansion project. *Rev. Sci. Instrum.* **80** (2009) 023101.

[11] Mildner, D.F.R. and Cook, J.C. Curved-straight neutron guide system with uniform spatial intensity distribution. *Nucl. Instrum. Meth.* A 592 (2008) 414.

[12] Copley, J.R.D. Transmission properties of short curved neutron guides: Part I. Acceptance diagram analysis and calculations, *Nucl. Instrum. Meth.* A **355** (1995) 469.

[13] Caglioti, G., Paoletti, A., and Ricci, F. Choice of collimators for a crystal spectrometer for neutron diffraction. *Nucl. Instr. Meth. Phys. Res.*, **3** (1958) 223-228.

[14] http://www.mcstas.org/download/components/sources/Source_gen.html (accessed 04 Feb. 20).

[15] Avdeev, M. and Hester, J.R. ECHIDNA: a decade of high-resolution neutron powder diffraction at OPAL. *J. Appl. Crystallogr.* **51** (2018) 1597-1604.

[16] Hewat, A. Design for a conventional high-resolution neutron powder diffractometer. *Nucl. Instr. Meth. Phys. Res.*, **127** (1975) 361-370.

[17] Wood, K. *et al*. QUOKKA, the pinhole small-angle neutron scattering instrument at the OPAL Research Reactor, Australia: design, performance, operation and scientific highlights. *J. Appl. Crystallogr.* **51** (2018) 294-314.

[18] Sokolova A. *et al*. Performance and characteristics of the BILBY time-of-flight small-angle neutron scattering instrument. *J. Appl. Crystallogr*. **52** (2019) 1-12.

[19] Dewhurst, C. D33 - a third small-angle neutron scattering instrument at the Institut Laue Langevin, *Meas. Sci. Technol.* **19**, 034007 (2008).